\documentclass[review]{elsarticle}

\usepackage{lineno,hyperref}
\usepackage{amsmath,amssymb}
\usepackage{booktabs}
\usepackage{color}
\usepackage{changepage}
\usepackage{diagbox}
\usepackage{makecell}
\modulolinenumbers[5]

\journal{Journal of \LaTeX\ Templates}

\begin{document}

\begin{frontmatter}

\title{Using an agent-based sexual-network model to analyze the impact of mitigation efforts for controlling chlamydia}

\author[mymainaddress1]{Asma Azizi\corref{mycorrespondingauthor}}
\cortext[mycorrespondingauthor]{Corresponding author}
\ead{aazizibo@asu.edu}
\address[mymainaddress1]{School of Human Evolution and Social Change; Simon A. Levin Mathematical Computational Modeling Science Center, Arizona State University, Tempe, AZ 85281}

\author[mysecondaryaddress2]{Jeremy Dewar}
\address[mysecondaryaddress2]{ Department of Mathematics, Tulane University, New Orleans, LA, USA, 70118}

\author[mysecondaryaddress3]{Zhuolin Qu}
\address[mysecondaryaddress3]{ Department of Mathematics
The University of Texas at San Antonio, San Antonio, Texas 78249}

\author[mysecondaryaddress2]{James Mac Hyman}


\begin{abstract}
~\\
Chlamydia trachomatis (Ct) is the most  reported sexually transmitted infection in the United States with a major cause of infertility, pelvic inflammatory disease, and ectopic pregnancy among women. Despite decades of screening women for Ct, rates  increase among young African Americans (AA). We create and analyze an  heterosexual agent-based network model to help understand the spread of Ct. We calibrate the model parameters to agree with  survey data showing Ct prevalence of $12\%$ of the women and  $10\%$ of the men in the $15-25$ year-old AA in New Orleans, Louisiana. Our model accounts for both long-term and casual  partnerships. The network captures the assortative mixing of individuals by preserving the joint-degree  distributions observed in the  data. We  compare the effectiveness of intervention strategies based on randomly screening men, notifying partners of infected people, which includes partner treatment, partner screening, and rescreening for infection.
We compare the difference between treating partners of an infected person both with and without testing them. We observe that although increased Ct screening, rescreening and treating most of the partners of infected people will reduce the prevalence, these mitigations alone are not sufficient to control the epidemic.  The current practice is to treat the partners of an infected individual, without first testing them for infection. The model predicts that if a sufficient number of the partners of all infected people are tested and treated, then there is a threshold condition where the epidemic can be mitigated. This threshold results from the expanded treatment network created by treating the partners of the infected partners of an individual. Although these conclusions can help design future Ct mitigation studies, we caution the reader that these conclusions are for the mathematical model, not the real world, and are contingent on the validity of the model assumptions.
 
\end{abstract}


\end{frontmatter}

\section{Introduction}

Chlamydia trachomatis (Ct) is the most commonly \textcolor{black}{notified bacterial} sexually transmitted infection (STI) in the United States, with over $1.8$ million cases each year \cite{torrone2014prevalence}.
It is a major cause of infertility, pelvic inflammatory disease (PID), and ectopic pregnancy among women \cite{cohen1998sexually, datta2012chlamydia,gottlieb2010introduction,gottlieb2010summary,hillis1996screening,lan1995chlamydia,pearlman1992review}, and has been associated with increased HIV acquisition \cite{cohen1998sexually,gottlieb2010introduction,lan1995chlamydia,pearlman1992review,hillis1996screening}.
Untreated, an estimated $14.8\%$ of women with Ct will develop PID \cite{price2013risk,price2016risk}, and $6\%$ will have tubal infertility \cite{lan1995chlamydia}. 
In southern US cities, including New Orleans, there is an ongoing epidemic of Ct in young African American (AA) adults.
A pilot study in this community \cite{kissinger2014check} found an average of $1.5$ sexual partners per person per three months, a relatively high turnover in sexual partners, and approximately $11\%$ prevalence of Ct infection.
The high prevalence and high turnover stress the need for more effective mitigation efforts to bring the epidemic under control. 

Further complicating the issue, some studies show that about $70\%-95\%$ of women and $90\%$ of men infected 
with Ct are asymptomatic and still transmit the infection to others \cite{farley2003asymptomatic,korenromp2002proportion}.
When Ct prevalence is high, regular screening is a practical approach to identify and treat infected individuals. 
\textcolor{black}{The US Preventive Services Task Force (USPSTF) recommends that sexually active women younger than $25$ years old, or older if they have multiple sexual partners, be screened for Ct as part of their physical exam \cite{lefevre2014screening}. However, untreated men may serve as a reservoir and reinfect treated women .} 
We investigate the impact \textcolor{black}{of} increased Ct screening of men in high prevalence areas on the disease prevalence. 
Typically, when someone is found to be infected, they are urged to encourage their sexual partners to be tested for infection. 
Sometimes the partners are treated without first being tested for infection.
If a partner is tested for infection and found to be infected, then their partners can be notified and treated, identifying a chain of high-risk individuals who might be spreading Ct. 

Transmission-based mathematical models can help the public health community to understand and to anticipate the spread of diseases in different populations and to evaluate the potential effectiveness of approaches for bringing the epidemic under the control \cite{azizi2016multi}.
These models create frameworks that capture the underlying Ct epidemiology and the heterosexual social structure underlying the transmission dynamics. Compartmental \cite{althaus2010transmission, de2008cost, de2006systematic, tuite2012estimation, townshend2000analysing, azizi2017risk, azizi2016multi, clarke2012exploring} and agent-based \cite{adams2007cost, andersen2006prediction,gillespie2012cost,roberts2007cost, low2007epidemiological, welte2005costs} mathematical models can help researchers to understand the transmission dynamics, and to analyze the efficiency and cost-benefit analysis of different intervention scenarios to control Ct infection in different regions. 

Althaus et al. \cite{althaus2010transmission} and Clarke et al. \cite{clarke2012exploring} used two different compartmental approaches to test the impact of screening programs on Ct infection, and both admitted that prevalence of Ct is not effectively sensitive to this program alone. Althaus et al. \cite{althaus2010transmission} identified the time to recovery from infection, and the duration of the asymptomatic period, as the two of the most essential model parameters governing the disease prevalence. When the underlying model is not static, Clarke et al. \cite{clarke2012exploring} demonstrated that random screening, if coupled with partner notification, is 
a cost-effective mitigation approach.

The efficiency of screening and partner notification strategies highly depends on the constructed model; that is, the result of one strategy can be different in the individual and population-level models \cite{althaus2012individual}.
Kretzschmar et al. \cite{kretzschmar1996modeling} used a stochastic network model based on pair formation and separation process to evaluate different screening and partner referral methodologies in controlling STIs such as Ct. Their results for Ct show that treating at least $50\%$ of partners of infected people can reduce the prevalence to a low level. They observed that the effectiveness of screening depends on the age, gender, and other characteristics of the targeted group.

The  heterosexual networks impacts the transmission dynamic and the effectiveness of different intervention approaches via capturing complex heterogeneous and biased mixing and sexual behavior of agents involved in the transmission process. 
Models must capture this underlying heterosexual network to predict how infections spread. 
Both the number of partners a person has (the degree distribution of the graph) and the number of partners their partners have (the joint-degree distribution of the graph) can impact this spread. 
We use estimates (provided in the supporting information Table (\ref{bjd1}) and (\ref{bjd2})) of both the heterosexual degree and joint-degree distribution from the ongoing New Orleans \textit{Check-it} of young AAs sexual behavior study \cite{kissinger2014check} to generate a heterosexual network that resembles sexual activity of this sub population of New Orleans.
That is, we use the B2K algorithm \cite{boroojeni2017generating} to preserve this joint-degree distribution and to generate an ensemble of heterosexual networks which reflects sexual partnership for our modeled New Orleans population.

The heterosexual partnerships are divided into long-term (primary) and short-term (casual) relationships. The casual partners change after a time, ranging from a few weeks to more than a year, while primary partners are maintained throughout the simulation. When changing casual partnerships, the degree and joint-degree distribution of this dynamic network are preserved. 

We model Ct transmission as a discrete-time Monte Carlo stochastic event on this dynamic network. 
The model is initialized to agree with the current New Orleans Ct prevalence.
We use sensitivity analysis to quantify the effectiveness of different prevention and intervention scenarios, including screening men, notification of partners, which includes partner treatment and partner screening (contact tracing), condom-use, and rescreening.

\section{Materials and Methods}

To be successful, a transmission model must accurately reflect what is happening in the real world, and the parameter values must be based on contemporary studies for the population being studied \cite{davies2014robust}.
The data used for our network was collected in $2016$ by two research studies in New Orleans, LA. One, a pilot study of community-based STI testing and treatment for AA men ages $15-25$ \cite{kissinger2014check} and the other, a study of an Internet-based unintended pregnancy prevention intervention for AA women ages $18-19$ \cite{green2014influence}. These two studies were reviewed and approved by the Tulane University Institutional Review Board. The $202$ men and $414$ women enrolled in these studies were asked for the number of different heterosexual partners they have had in the past three months (for women) or two months (for men). Women were also asked to estimate how many partners that their partners have had in the past two months.

The survey results were used to construct the bipartite heterosexual network of $P^m$ men and $P^w$ women. Using the \emph{Check-it} survey data for the number of partners, and the number of partners of partners in the last three months \cite{kissinger2014check,green2014influence}, we used the B2K algorithm \cite{boroojeni2017generating} to generate an ensemble of heterosexual networks for our model. Boroojeni et al. \cite{boroojeni2017generating} B2K algorithm preserves the degree (number of partners) and joint-degree (number of partners of partners) distribution of nodes (individuals) estimated from the survey data and provided in the supporting information.
The generated network for the sexually active population preserves the distribution for the number of partners that men and women have had in the past three months. The network also preserves the 
distribution for the number of partners of their partners (the joint-degree 
distribution) \cite{boroojeni2017generating}.

Each node \textbf{i} in the network represents a person, denoted by the index \textbf{i}, and each edge $\textbf{e}_\textbf{ij}$ represents sexual partnership between two nodes \textbf{i} and \textbf{j} on that day. 
The underlying network for the sexually active population is weighted, where the weight $0<w_\textbf{ij}\leq 1$ for edge $\textbf{e}_\textbf{ij}$ is \textcolor{black}{the probability, of sexual act between partners \textbf{i} and \textbf{j} on any specific day. The simulation network is dynamic and changes each day. For example, if $w_\textbf{ij}=1/7$, these two partners engage in a sexual act, on average, once a week, and the edge between nodes \textbf{i} and \textbf{j} is present, on average, once every seven days. This is implemented in the model as a stochastic process, the edge $\textbf{e}_\textbf{ij}$ will exist (turn on) with probability $w_\textbf{ij}$.}

The structure of the underlying weighted network changes every $T$ days when people change their casual partnerships. The updated underlying network with the new casual partners has the same degree and joint-degree distributions as the original network. We assume that every man has one primary partnership that is maintained for the entire simulation; all other partnerships are causal. When defining the initial network, we select the primary partner for each man as his female partner with the fewest number of partners, and within two years of his age.
To change the casual partners, we use the subgraph of primary partnership -as initial sexual network- and B2K algorithm in \cite{boroojeni2017generating} to generate a set of networks, and then every T days we update the network by randomly choosing one network from this set. 

In our stochastic Susceptible--Infectious--Susceptible (SIS) model, a person \textbf{i} is either infected with Ct, $I_\textbf{i}(t)$, or susceptible to being infected, $S_\textbf{i}(t)$.
During the day $t$, an infected person, $I_\textbf{j}(t)$, can infect any of their susceptible sexual partners, $S_\textbf{i}(t)$. 
We define $\lambda_{ij}$ as the probability that $S_\textbf{i}(t)$ will be infected by $I_\textbf{j}(t)$ by the end of the day, $S_i(t) \overset{\lambda_{ij}}{\rightarrow} I_i(t+1)$.
Similarly, we define $\gamma_j$ as the probability that an infected person, $I_\textbf{j}(t)$, will recover by the end of the day, $I_j (t) \overset{\gamma_j}{\rightarrow} S_j(t+1)$.

\subsection{Force of infection} 
The force of infection, $\lambda_{ij}(t)$, is the probability that a susceptible person $S_j$ is infected on day $t$ by $I_j$. The infection transmission depends on the probability of a sexual act between person $i$ and $j$ on a typical day, as defined by edge weight, $w_{ij}$, in the model. 
We define $\beta_{nc}$ as the probability of transmission per act when a condom is not used, and $\beta_{c}$ as the reduced probability of transmission per act when a condom is used.
The forces of infection between \textbf{i} and \textbf{j} for when condom is not used, $\lambda^{nc}_{ij}$, and for when condom is used, $\lambda^{c}_{ij}$, are defined by
\begin{equation}
\lambda^{nc}_{ij}=\begin{cases} 
\beta_{nc} & \textnormal{with probability}~w_{ij} \\
0 & \textnormal{with probability}~1-w_{ij} 
\end{cases}
,~~
\lambda^c_{ij}=\begin{cases} 
\beta_{c} & \textnormal{with~probability}~w_{ij} \\
0 & \textnormal{with~probability}~1-w_{ij}~~
\end{cases}.
\end{equation}
Consistent condom use approximately $98\%$ effective in preventing Ct transmission when used correctly \cite{warner2004condom,niccolai2005condom,paz2005effect,warner2006condom}.
This effectiveness is reduced to $85\%$ if the condom is incorrectly used \cite{condom}. 
There is very little quantitative data on correct condom use, and correlating consistent condom with long-term versus casual partners. In the model we assumed that condom use is $\epsilon=90\%$ effective in preventing the infection from being transmitted, that is, $\beta_c=(1-\epsilon)\beta_{nc}=0.1\beta_{nc}$. We assume that, the effectiveness of condom is independent of gender of donors and recipient, that is, if $\beta^{m2w}$ and $\beta^{w2m}$ are defined as probability of transmission from men to women and from women to men, then in the case of condom use these values will reduce by the same factor $(1-\epsilon)$. We also assume that condoms are used more often in casual (riskier) partnerships ($\kappa$ of the time) than in long-term partnerships.

\subsection{Recovery from infection}
The model accounts for infected people recovering through natural recovery or after being treated with antibiotics. We assume that a fraction of the people treated for infection returns to be retested to see if they have been reinfected. We also assume that all infected people eventually recover and return to susceptible status, even when they have not been treated. 
In the model, the time for \textbf{natural (untreated) recovery} has an exponential distribution with an average time of infection of \textcolor{black}{$\tau_n=1/\gamma_n$} days, and the duration of infection for an individual is a random number from this distribution.

However, some infected individuals become recovered through treatment which is part on intervention program, explained in details in the next subsection.

\subsection{Intervention strategies}
We assume that the treatment is $100\%$, and the time to recover after treatment is a log-normal distribution with the parameters of $\tau^t=1/\gamma^t$ day and $\sigma^2=0.25$. 
That is, the duration of infection for a treated infected person \textbf{k} is defined by a random number chosen from a log-normal distribution, $\log \mathcal{N}(\tau^t, 0.25)$, rounded to the nearest day. 
In the model, if that number of days is smaller than the duration remaining for naturally clearing the disease, then the shorter time is used for the recovery period. 

Each year, a fraction of the population is tested for Ct infection through a routine medical exam (random screening), after observing symptoms, or after being notified that one of their previous partners was infected. 

\sloppy\emph{\textbf{Random Screening}}: We define random screening as testing for infection when there are no compelling reasons to suspect a person is infected. For example, random screening might be part of a routine physical exam and is an effective mitigation policy to identify asymptomatic infections.
We assume that the fraction $\sigma_y\%$ of people are randomly screened each year. Each individual is screened with probability $\sigma_d=1-(1-\sigma_y)^{\frac{1}{365}}\%$ each day. \textcolor{black}{It is relatively rare for the Ct screening test to give a false negative result, and in the model, we assume that screening is 100\% accurate. } 

\emph{\textbf{ Notification of Partners}}: 
We assume that an infected person encourages some of their partners to be treated or tested for infection. We define
$\theta_n$ as the fraction of their partners who notified, which  $\theta_t$ fraction of these notified partners follow treatment, and the rest ($\theta_s=1-\theta_t$ fraction \textcolor{black}{ of notified partners}) \textcolor{black}{test for the infection}.

Therefore,  a \emph{notified partner} is the partner of an infected person who seeks treatment or testing as a direct result of the screening. We can divide all partners  of  person as follow: 
\begin{enumerate}[(1)]
\item Partner Treatment: $\theta_n\theta_t$ fraction of all partners that are notified and  treated, without first testing for infection.
\item Partner Screening: $\theta_n\theta_s$ fraction of all partners are notified and screened for infection and then start treatment if they are infected. 
\item Do nothing: $1-\theta_n=1-\theta_n\theta_t-\theta_n\theta_s$ fraction are neither tested nor  treated.
\end{enumerate}

The model includes a time-lag of $\tau_N$ days between the day a person is found to be infected, and the day their partners are notified and take action.

\textit{\textbf{Rescreening}}:
A common practice in disease control is rescreening.
People found to be infected are treated and asked to return after a short period to be tested again for infection.
We assume that a fraction, $\sigma_r$, of the treated people return for retesting $\tau_R$ days after treatment.

\subsection{Model initialization}
We initialize our model based on estimates for the current Ct epidemic in New Orleans at the time the Ct prevalence was $i_0$.
The initially infected people are not randomly distributed in an otherwise susceptible population. They are distributed as they would be as part of an emerging epidemic that started sometime in the past.
We call these initial conditions \textit{balanced} because when the simulation starts, the infected and susceptible populations, along with the duration of infection, are in balance with the distributions for an emerging epidemic \cite{hyman2001initialization}.
When the initial conditions are not balanced, then there is usually a rapid (nonphysical) initial transient of infections that quickly dies out as the infected and susceptible populations relax to a realistic infection network.

\begin{figure}[tb]
\centering
\includegraphics[width=100mm,scale=0.5]{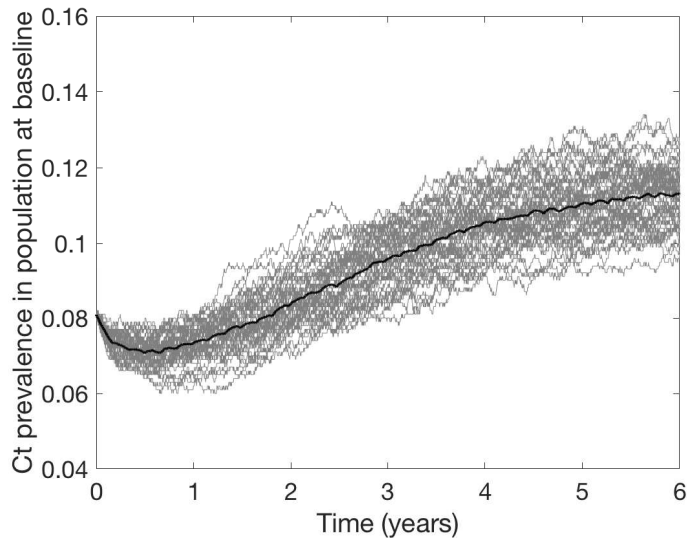}
\caption{\textbf{Prevalence increases to reach the current quasi-stationary state:} About  $11\%$ of total population are infected at the quasi-stationary state for the baseline model parameters. The standard deviation of the prevalence is approximately 1\%. This is in good agreement with the current prevalence in New Orleans 15-25 year-old AA population. The light areas are the result for $50$ different stochastic simulations and the dark curve is the mean value of those simulations. }
\label{forbeta}
\end{figure}

To define the balanced initial conditions, we start an epidemic in the past by randomly infecting a few high degree individuals.
We then advance the simulation until the epidemic grows to the prevalence of $i_0$.
We then reset the time clock to zero and use this distribution of infected people, complete with their current infection timetable, as our initial conditions. 
Because these are stochastic simulations, when doing an ensemble of simulations, we reinitialize each simulation by seeding different initial infected individuals. 
Fig \ref{forbeta} illustrates the typical progression of the epidemic to reach the current Ct prevalence of $10\%$ in sexually active men and $12\%$ women in the $15-25$ year-old New Orleans AA community \cite{kissinger2014check}. These estimates are based on the current \emph{Check it} survey data of approximately $1084$ AAs in New Orleans and have a standard error of about $1\%$. 

The numerical simulations comparing the different mitigation strategies all start at this endemic stochastic equilibrium.

\section{Result}

Each simulation is a single realization of a stochastic epidemic on one of an ensemble of heterosexual networks that share the same joint-degree distribution. In our investigations, we have found that the qualitative trends for the predictions as a function of these assumptions are insensitive to the details of networks with the same joint-degree distribution. 
We compared these trends for the predicted impact of increased random screening, infected partner treatment and screening, and rescreening of treated individuals on the prevalence of Ct infection.
All of the simulations start at a balanced equilibrium obtained with the model baseline parameters in Table \ref{paras}, unless stated otherwise.

In screening intervention, we assume that the day of screening is the day that an infected person found through screening starts treatment, that is, getting the result and starting treatment happens on the same day. We assume that a treated person does not change his/her behavior. That is the treated person regular sexual acts with partners and may infect susceptible partners during the treatment period with the same probability of transmission for when natural recovery. Once started, the interventions maintained as the solution approaches a new quasi-stationary state corresponding to implemented intervention. 
\begin{table}[t]
\begin{adjustwidth}{-0.in}{0in} 
\centering
\caption{\textbf{Parameters definition and value:} the model parameters describing the transmission of Ct infection, as well as recovery associated with natural recovery, screening, and notification of partners were obtained from the literature \cite{kissinger2014check, heijne2011role,althaus2010transmission,morre1998monitoring}, but other parameters are calibrated to biological, behavioral, and epidemiological data from general heterosexual population resides in New Orleans.
Probability of transmission per act is calibrated to a baseline prevalence of $12\%$ among women and $10\%$ among men.
}

\resizebox{\columnwidth}{!}{\begin{tabular}{p{.9in}p{3.24in}lccc}
\toprule[1.5pt]
\textbf{Parameter} & \textbf{Description} & \textbf{Baseline} & \textbf{Unit} & Reference \\[0.50ex]
\hline
$\Delta t$ & Time step & 1 & day&-- \\[0.50ex]
\hline
$P^m$($P^w)$ & Population of men (women) & $2000(3000)$ & people & Assumed \\
\hline
$\beta^{m2w}$& Probability of transmission per act from men to women & $0.10$ & -- &Calibrated\\
\hline
$\beta^{w2m}$& Probability of transmission per act from women to men & $0.04$ & -- & Calibrated \\
\hline
$\kappa$& Fraction of times that condoms are used during sex & $0.58$ & -- &\cite{kissinger2014check} \\
\hline
\textcolor{black}{$\epsilon$}& \textcolor{black}{Condom effectiveness} &\textcolor{black}{$0.90$} & -- & \textcolor{black}{Assumed}\\
\hline
$1/\gamma^n$& Average time to recover without treatment & $365$ & days &\cite{molano2005natural,althaus2010transmission} \\
\hline 
$1/\gamma^t$& Average time to recover with treatment & $7$ & days &\cite{morre1998monitoring} \\
\hline
$\sigma^m_y$& Fraction of men randomly screened per year & $0.05$ & -- &\cite{kissinger2014check} \\ 
\hline
$\sigma^w_y$& Fraction of women randomly screened per year & $0.45$ & -- &\cite{kissinger2014check} \\
\hline
$\sigma_r$& Fraction of infected people return for rescreening & $0.10$ & -- &Assumed \\ 
\hline
Sensitivity& Screening sensitivity & \textcolor{black}{$100\%$} & -- &\textcolor{black}{Assumed} \\ 
\hline
$\theta_n$& Fraction of the partners of an infected person who are notified and do test or treated for infection & $0.26$ & -- &\cite{kissinger2014check}\\
\hline
$\theta_t$& Fraction of notified partners of an infected person who are treated without testing & $0.75$ & -- &\cite{kissinger2014check} \\
\hline
$\theta_s$& Fraction of notified partners of an infected person who are tested and treated for infection & $0.25$ & -- &\cite{kissinger2014check} \\
\hline
$\tau_N$ & Time lag of partner notification & $5$ & days & Assumed \\
\hline
$\tau_R$ & Time lag of re-screening & $100$ & days & Calculated \\
\hline
$T$ & Time period for casual partner change & $60$ & days & Assumed \\
\bottomrule[1.5pt]
\end{tabular}}
\label{paras}
\end{adjustwidth}
\end{table}

\subsection{Sensitivity analysis}
Table \ref{SA} gives the relative sensitivity indices for the change in the mean values of the prevalence \pmb{$Pr$} with respect to the model parameters at the baseline parameter values, $\mathbb{S}^{Pr}_p$. If the parameter $P$ changes by $x\%$, then \pmb{$Pr$} will change by $x\mathbb{S}^{Pr}_p\%$.

The prevalence is most sensitive to the fraction of time condom used, $\kappa$, closely followed by uncertainty in the estimate of infection period in the absence of treatment, $1/\gamma^n$, and then the probability of transmission per sexual act, $\beta^{m2w}$ and $\beta^{w2m}$. 
Variation in other model parameters had a minor impact on the model outcomes in the  sensitivity analysis ($\leq 1\%$ ).

\begin{table}[htp]
\centering
\begin{tabular}{cccc|ccccc}
\toprule
& \multicolumn{7}{c}{\textbf{Relative sensitivity index of prevalence \pmb{$Pr$}}}
\\
\cmidrule(rl){2-5} \cmidrule(rl){5-8}
& {Parameter p} & {Baseline } & {$\mathbb{S}^{Pr}_p$}
& {Parameter p} & {Baseline } & {$\mathbb{S}^{Pr}_p$}&\\
\midrule
& $\kappa$ & $0.58$ & $-2.47$ & $1/\gamma^n$ & $365$ & $1.66$&\\
& $\beta^{w2m}$ & $0.04$ & $1.03$ & $\beta^{m2w}$ & $0.10$ & $1.02$&\\
& $1/\gamma^t$ & $7$ & $-0.15$ & $\sigma_r$ & $0.10$ & $-0.02$&\\ 
& $\tau_R$ & $100$ & $-0.01$ & $\tau_N$ & $5$ & $-0.0569$ &\\
\bottomrule
\end{tabular}
\caption{The sensitivity indexes of prevalence at quasi-steady state, {$Pr$}, with respect to parameters of the model, at the baseline parameter values where $Pr=0.11$. The most sensitive parameter for the prevalence is the fraction of sexual acts condom used, $\kappa$. If condom use increased by $10\%$, then the prevalence, $Pr$, would decrease by $-24.7\% = -2.47 \times 10\%$. The second most sensitive parameter is the natural recovery period, $\tau^n=1/\gamma^n$.}
\label{SA}
\end{table}

\subsection{Random screening} 
\begin{figure}[tb]
\centering
\includegraphics[scale=0.4]{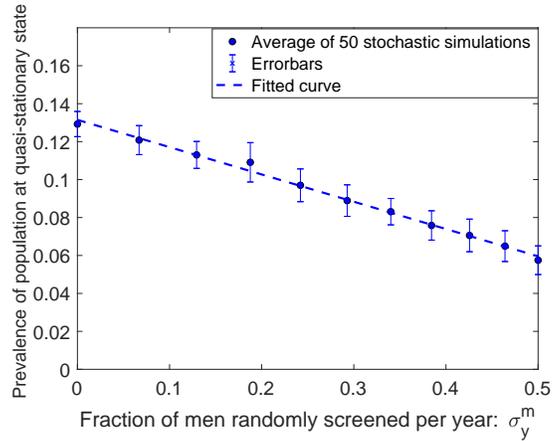}
\caption{\textbf{\textcolor{black}{Quasi-stationary} state prevalence of the population versus the fraction of men are screened each year} (All other parameters are defined as in Table \ref{paras}): the circles are the average of $50$ different stochastic simulations, and the error bars are $95\%$ confidence intervals. Screening men randomly by $50\%$ reduces prevalence by $7\%$ ($SI=3.5\%$), which is not effective enough to implement as a sole intervention. }
\label{just_screening}
\end{figure}

To determine the effectiveness of increasing the number of men screened for Ct per year, we compare the \textcolor{black}{quasi-stationary} state prevalence by varying the fraction of men who are randomly screened each year, $\sigma^m_y$.
The current screening rate for young men for Ct in high prevalence areas, like New Orleans, is low. This scenario can estimate the cost-effectiveness of increased screening of young men on the Ct prevalence in women \cite{gift2008program, gopalappa2013cost}.
Fig \ref{just_screening} shows a reduction in the overall Ct prevalence as the number of men randomly screened for Ct increases from $0$ to $50\%$, $0\leq \sigma^m_y\leq 0.5$.
The least-square linear fit suggests that the \textcolor{black}{quasi-stationary} state Ct prevalence will decrease by $1.4\%$ for every additional $10\%$ of the men screened during a year ($SI=6.5\%$).
Though a drop of seven percent in prevalence is an admirable decrease, increased screening alone would not be sufficient to control Ct. In this simulation, the casual partners of men change every $60$ days.

\subsection{Notification of partners} 
\textcolor{black}{The scenario of notifying partners quantifies the impact of giving an infected person's partners a chance to be tested and treated. As we mentioned before, the fraction $\theta_n$ of an infected person's partners are notified about infection. We assume that some of these partners are treated, and others are screened for infection.}

\subsubsection{Partner treatment only}

\begin{figure}[tb]
\centering
\includegraphics[width=0.7\linewidth]{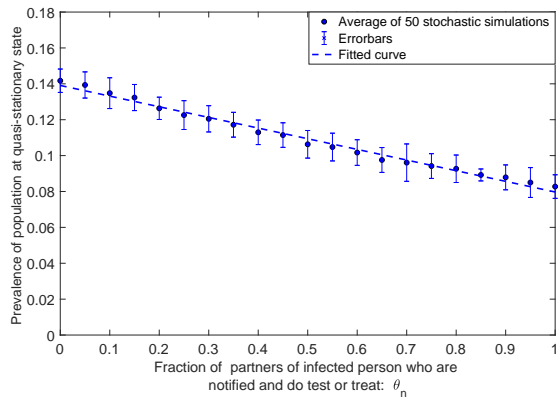}
\caption{\textbf{\textcolor{black}{Quasi-stationary} state prevalence of the population versus the fraction of partners are notified for partner treatment only scenario ($\pmb{\theta_t=1}$)}: The circles are the mean of $50$ different stochastic simulations and error bars are 95\% confidence intervals. Partner treatment is only mildly effective, and the prevalence remains high ($8\%$), even when all the partners of treated people are treated.}
\label{pt}
\end{figure}

In partner-treatment, we assume when someone is found to be infected, then a fraction $\theta_n$ of their partners are notified, and then all of the notified partners seek treatment without testing; that is, $\theta_t=1$.
The Fig \ref{pt} shows the impact of partner treatment for different values of partner notification values changing from  $\theta_n=0$ to $\theta_n=1$.
The least-square linear fit suggests that the \textcolor{black}{quasi-stationary} state Ct prevalence decreases by $0.07$ for every $10\%$ increment in the fraction of notified partners seeking treatment. 
In this simulation, the casual partners of men change every $60$ days.

\subsubsection{Partner screening only}
\begin{figure}[tb]
\centering
\includegraphics[width=0.7\linewidth]{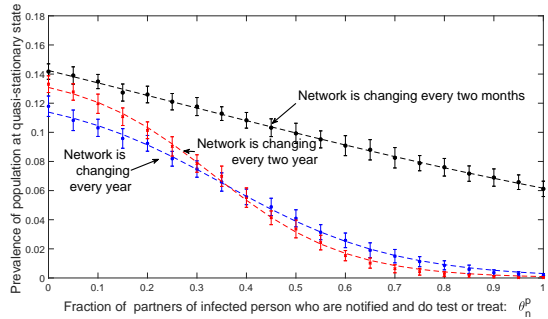}
\caption{\textbf{\textcolor{black}{Quasi-stationary} state prevalence of the population versus the fraction of partners are notified for partner screening only scenario ($\pmb{\theta_s=1}$)}: All other parameters are at the baseline values in Table \ref{paras}). The figure illustrates how the effectiveness of partner notification increases as casual partners are changed less often from every $60$ days (black line), every year (blue curve), and every two years (red curve). The circles are the mean of $50$ different stochastic simulations and error bars are $95\%$ confidence intervals. Partner screening is more effective in situations where the casual partners less often.} 
\label{ptt1}
\end{figure}

\begin{figure}[tb]
\centering
\includegraphics[width=0.7\linewidth]{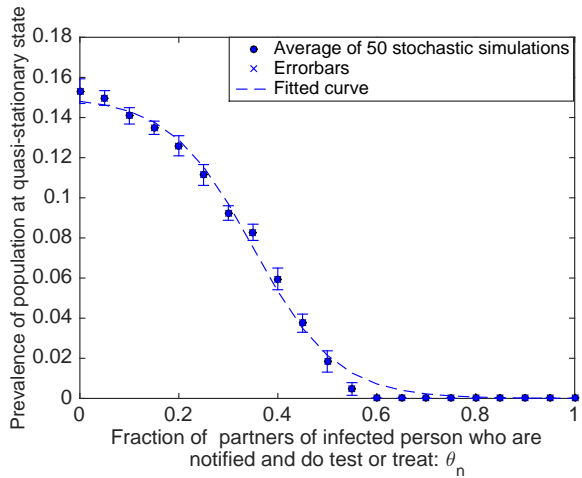}
\caption{\textbf{Quasi-stationary state prevalence of the population versus the fraction of partners are notified when all notified partners first follow screening and treatment if infected i.e} $\pmb{\theta_s=1}$\textbf{ for static network} (All other parameters are defined as in Table \ref{paras}): the circles are the mean of $50$ different stochastic simulations and error bars are $95\%$ confidence intervals. Partner screening approach is highly effective when $\theta_n$ is big enough, that is, when $\theta_n\geq 0.4$ and $\theta_s=1$, the Ct prevalence rapidly decays to zero.}
\label{ptt2}
\end{figure}

To quantify the impact of screening the partners of an infected person, where partners are tested and then treated if they are infected. We assume that all notified partners of an infected person are screened; that is, $\theta_s=1$.
Fig \ref{ptt1} shows the impact as $\theta_{n}$ varies from $0$ to $1$ for the cases where the casual partners of the men change, on average, every $60$ days (black line), every year (blue curve), and every two years (red curve), and Fig \ref{ptt2} is for the case men do not change their casual partners (static network). The simulations predict the test and treat partner notification is most effective when casual partnerships are longer term. This is expected because when a partner is found to be infected, and this partner's other (long-term) partners are tested, then the contact tracing is a branching process and is more likely to identify and treat the underlying infected sexual network. The resulting nonlinear effect is evident in the logistic-shaped curve $\theta_n$ when the casual-partners change less often.

\subsubsection*{Partner treatment and screening}
Some of the notified partners will seek treatment without testing, and some will allow themselves to be tested before being treated.
We quantify the effectiveness of this mixture of the two previous scenarios. We let the fraction of notified  partners, $\theta_n$, range between $0.10$ and $0.80$. We also vary the fraction of these notified partners who seek just treatment, $\theta_{t}$. In these simulations, $\theta_s=1-\theta_t$ of the population is screened for infection each year, and the casual partners are updated every two years.

\begin{figure}[tb]
\centering
\includegraphics[width=0.7\linewidth]{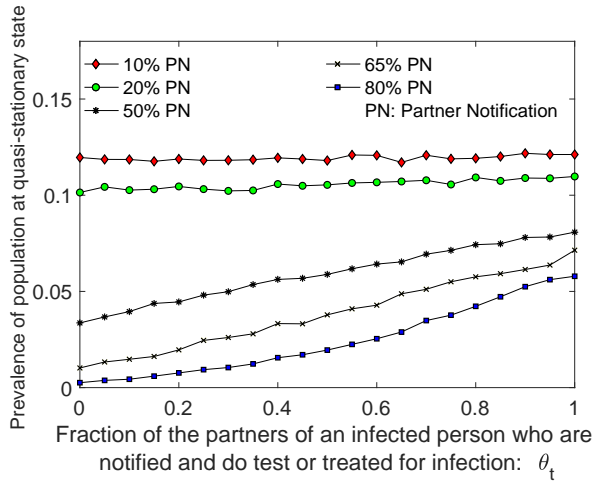}
\caption{ \textcolor{black}{\textbf{Prevalence at quasi-stationary state increases when the fraction of partners notified are treated and not tested} (All other parameters are defined as in Table \ref{paras}): every curve is the mean of $50$ different stochastic simulations. When only a few partners of an infected person are notified, $\theta_n$ is small, then partner treatment and partner screening have a similar small impact on Ct prevalence. When more partners of infected people take action, $\theta_n$ increases, then the partner screening strategy is more effective in controlling the infection.}
}
\label{pt_ptt}
\end{figure}

When few partners are notified and take action ($\theta_n$ is small), then partner treatment and partner screening have almost the same impact on controlling the prevalence.
For example, for $\theta_n=0.10, 0.20$, the prevalence versus $\theta_t=1-\theta_s$ is flat; that is, there is no difference between cases if partners follow treatment without testing or first test and then treat if infected, Fig \ref{pt_ptt}.

As $\theta_n$ increases the partner screening becomes a highly successful mitigation policy.
Consider the case when half of the partners are notified and take action, $\theta_n =0.5$, and half of them are screened for infection, $\theta_s=0.5$, and the other half are treated without testing, $\theta_{t}=0.5$.
That is, half of an infected person's partners do nothing, the fraction $\theta_n\theta_t=0.5\times0.5=0.25$ are treated without testing for infection, and the fraction $\theta_n\theta_s=0.5\times0.5=0.25$ are tested and treated if found infected.
If any of the tested notified partners of the infected person are found to be infected, their partners are then notified, and the cycle repeats to spread out and identify more infected people. 
This conditional percolation of screening through the sexual network is why this policy is so effective.
In this case, the prevalence reduction is $6\%$. Thus when compared with all notified partners following treatment without testing, $\theta_{t}=1$, which reduces the prevalence by only $1\%$, conditional percolation is more effective. But compared to all notified partners following test and treat if necessary, $\theta_s=1$, which reduces the prevalence by $11\%$, this combined scenario is not the one to select, Fig \ref{pt_ptt}. 

\subsection{Rescreening} 
The rescreening scenario quantifies the effectiveness of different time-lags for rescreening and quantifies the impact of rescreening on the prevalence of Ct. 
\subsubsection*{Interval for rescreening}
People who are found to be infected are more likely to be reinfected in the future.
Repeated Ct infection can be the result of sexual activity with a new partner, or being reinfected from an existing infected partner.
It makes sense to ask the infected people who were treated to return in a few months for retesting.
We use the model to compare the rates of reinfection to help optimize the time, $\tau_r$, from treatment to rescreening.

The time $\tau_r$ between treatment and rescreening should be long enough so that it is likely that the person would be reinfected if one of their partners is still infected. If $\tau_r$ is too long, then a reinfected person could infect others. The current CDC guidelines recommend that people are rescreened for infection three months after treatment \cite{peterman2006high}. 
Past studies have observed that about $25\%$ of the rescreened people are again found to be infected by three months.

\begin{figure}[tb]
\centering
\includegraphics[scale=0.35]{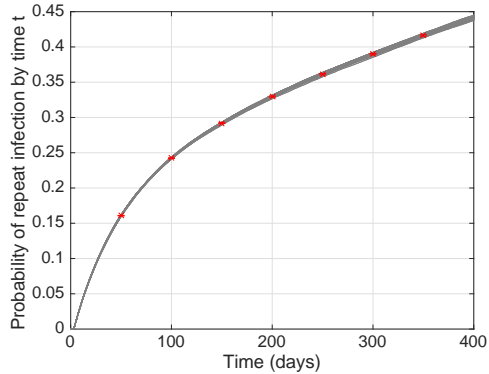}
\caption{\textbf{Truncated cumulative probability distribution of time between treatment and reinfection with Ct}: fifty different curves from $50$ stochastic simulations and $95\%$ confidence interval are shown in this figure.
About $25\%$ of the treated people are again infected after $100$ days. This increases to about $45\%$ are reinfected after almost a year. }
\label{cdf-rescreening}
\end{figure}

We plot the cumulative distribution of time between screening and reinfection events in Fig \ref{cdf-rescreening}.
The Figure demonstrates that our model also predicts that about $25\%$ of treated individuals are again infected after $100$ days.
The model predicts that the Ct prevalence of the treated population exceeds the prevalence for the whole population after two months and suggests that the CDC guidelines for rescreening could be shortened slightly. 

\subsubsection{Rescreening rate}
\begin{figure}[tb]
\centering
\hspace{1cm}
\includegraphics[scale=0.3]{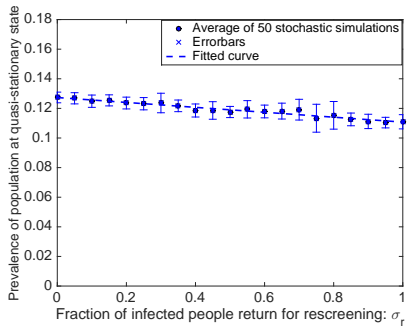}
\caption{\textbf{\textcolor{black}{Quasi-stationary} state prevalence of population versus the fraction of treated people who come back for screening, $\pmb{\sigma_r}$} (All other parameters are defined as in Table \ref{paras}): the circles are the mean of $50$ different stochastic simulations and error bars are $95\%$ confidence intervals. Rescreening all the infected people reduces the prevalence by $2\%$.
}
\label{rescreening}
\end{figure}

Typically, only about $10\%$ ($\sigma_r =0.1$) of screened individuals return for rescreening.
To understand if increasing the rate that people return for rescreening would have a significant impact on Ct prevalence, we varied the fraction of treated individuals who return for rescreening at $100$ days after being treated.
The Fig \ref{rescreening} quantifies the prevalence of Ct at quasi-stationary state dependent on rescreening rate $\sigma_r$: there is a negative correlation between prevalence at quasi-stationary state and $\sigma_r$ when $\sigma_r$ fraction of screened individuals returns for rescreening, if $\sigma_r$ fraction of screened individuals follow screening again then the prevalence reduces roughly by $0.02\sigma_r$ ($SI=2\%$). We observe that this rescreening result is insensitive to partner change in the network.

\section{Discussion}
Using a heterosexual behavior survey and Ct prevalence data for the sexually active young AA population in New Orleans, we  created an agent-based dynamic  network model to understand how the infection spreads and what mitigation approaches can slow it down.
In the model, men and women are represented by the network nodes, and the sexual partners are characterized by edges between the nodes. 
The edges between partners in the network dynamically appear and disappear each day, depending on if the individuals engage in sex on that day. The partners of men are divided into primary and casual ones, and their casual partners are changing over time.
One novel property of our model is that the joint-degree distribution of the network- which captures the correlation of an individual's risk (their number of partners) with their partner's risk (number of partners of their partners)- is preserved while changing casual partners. That is, in whole time frame of the simulation and in spite of changing the sexual network, we are capturing the information collected from data.

There is a wide range of reported values for the probability of transmission per sexual contact, ranging from from $0.04$ to $0.16$ \cite{low2007epidemiological,welte2005costs,adams2007cost,andersen2006prediction,gillespie2012cost,roberts2007cost}. We calibrated this parameter to the current prevalence of Ct among adolescents and young adult AAs in New Orleans \cite{kissinger2014check}. In our model, we defined the probability of transmission from man to woman as $0.1$ per contact, and from woman to man is $0.04$ per contact.

We use this model to quantify the impact of increasing screening of men for infection, notification of partners, and rescreening of treated individuals on reducing Ct prevalence. 
We observed that increasing Ct screening of men has a modest impact on reducing Ct prevalence in the young adult AAs in New Orleans, Fig \ref{just_screening}.
Starting at a baseline of $11\%$ prevalence under the assumption that $45\%$ of the women are being screened each year for Ct, then increasing the screening of men from $0\%$ to $50\%$ reduces the overall Ct prevalence to $8\%$.
Our observation that partner positivity is insensitive to screening is consistent with previous studies \cite{clarke2012exploring}. 


In evaluating the effectiveness of partner-notification, we assumed some of the partners of an infected person would seek treatment (without testing) or be screened (tested and treated) for infection.
We observed that if most of the notified partners are treated, without testing, then this mitigation has only a modest impact on Ct prevalence.
This practice, although common in disease control today, is not as effective as partner screening.
When individuals change their partner less frequently, and the partners of an infected person were tested before treatment, there was a tipping point where partner screening would bring the epidemic under control.
For example, when casual partners do not change very often, then when over $40\%$ of notified partners of all the infected people are screened for infection, the Ct prevalence rapidly decreased to very low levels, Fig \ref{ptt2}.
This critical threshold represents the partner screening level, where a contact tracing tree can spread through the heterosexual network to identify and treat most of the infected people. 
Our model indicates that this is by far the most effective approach for bringing the epidemic under control.

However, partner screening is more expensive than partner treatment.
The partner treatment and screening suggests that when the fraction of partners took action ($\theta_n$ is small), then partner screening may not be a good strategy compared to partner treatment. 
But if a large enough fraction of partners is notified, then it is better to test and treat (partner screening) to control the spread of Ct effectively. These results of the impact of notification of partner are close to results from \cite{kretzschmar1996modeling} who found for
Ct, contact tracing is less effective at lower
percentages when partners are treated, but with increasing levels of contact tracing it will be 
a highly effective intervention strategy.


In rescreening, infected individuals return for testing a few months after being treated.
We used the model to estimate the probability that a treated person would be reinfected as a function of the time since they were treated. 
The CDC guidelines recommend that treated people return for screening three months after treatment. 
We observed that for the case of $13\%$ infected population, about $25\%$ of the treated population were reinfected three months after treatment. 
Previously infected people are more likely to be infected again than the average person. Although the rescreening has only a small impact on the overall Ct prevalence, it is an effective way of identifying reinfection.
Even though there is a high chance of reinfection when the individual's behavior does not change, we do not observe an effective impact on the prevalence of Ct by monitoring infected individuals.
The rescreening program has a trend similar to screening, and none of them are effective as sole intervention because they are not able to find the chain of infection like partner screening.
On the other hand, the sensitivity of prevalence to rescreening is less than that of screening, indicating the fact that for a limited budget, the idea of finding more people to screen, random screening, is more effective than frequent screening for fewer people.

Although our model takes into account different important factors for STI transmission and can be used to assess the relative impact of different mitigation, it is still too simplistic to be used for quantitative predictions. 
One of the most important behaviors to capture in an STI model is how often condoms are used in long-term and casual partnerships. 
Although our model includes condom use, it does not account for behavior changes, such as increased condom use after being treated for the infection. 
Our future research will improve the model so we can better quantify the impact of counseling and behavioral changes such as increasing condom use or partner notification rates. We are expanding our data analysis to include a cost-benefit analysis and estimate the averted PID cases in women.

Our bipartite heterosexual network model was constructed based on the correlations between the number of partners a person has and the number of partners their partners have. For our future work, we will extend our assortative mixing model to improve our assumptions where sexual partnerships are better characterized by their ages, ethnicity, social groups, economic status, and geographic location. 
We will focus on validating the model predictions and identifying which trends and quantities can and cannot be predicted within limits of the model uncertainty. 
Our preliminary studies indicate that the qualitative findings of this paper are relatively insensitive to adding these additional mixing constraints. 
Although the model is still too simplest to directly guide mitigation efforts, the qualitative trends predicted by these simulations can be useful in designing studies to quantify the effectiveness of different mitigation efforts.

\section*{Supporting information }

{\color{black}
Tables \ref{bjd1} and \ref{bjd2} define the joint-degree distribution for the number of partners of partners for men and women in the population used in this study. The element $p_{ij}$ of Table \ref{bjd2} is the fraction of edges between men with $i$ partners and women with $j$ partners. The fraction of men with $i$ partners is the sum of the $i^{th}$ row divided by $i$, and the fraction of women with $j$ partners is the sum of the $j^{th}$ column divided by value $j$. The algorithm in \cite{boroojeni2017generating} guarantees that the joint-degree distribution of generated network is consistent with the Kissinger et al. \emph{Check It} study of the behavior of $15-25$ year-old sexually active African American men and women in New Orleans \cite{kissinger2014check}, Table \ref{bjd1}.

\begin{table}[htp]
\begin {center}
\caption{Joint-degree Table of sample data for women participant in the last three months.}
\begin{tabular}{>{\color{black}}c>{\color{black}}c>{\color{black}}c >{\color{black}}c >{\color{black}}c >{\color{black}}c >{\color{black}}c >{\color{black}}c >{\color{black}}c >{\color{black}}c }
\hline
 \multicolumn{7}{c}{\textcolor{black}{Number of Partners for Women Participants}}\\ 
\hline
& & 1 & 2 & 3 & 4 & 5 & 6 \\ 
\hline
  & 1 & 197 & 46 & 16 & 4 & 0 & 3 \\
 & 2 & 56 & 34 & 11 & 2 & 2 & 0\\
 & 3 & 28 & 23 & 3 & 4 & 3 & 5 \\ 
 & 4 & 9 & 30 & 10 & 0 & 0 & 1 \\ 
 & 5 & 2 & 2 & 2 & 0 & 0 & 0 \\ 
Number of& 6 & 3 & 1 & 1 & 1 & 4&0 \\ 
Partners & 7 & 0 & 2 & 1 & 0 & 0 &1 \\ 
for & 8 & 1 & 0 & 0 & 0 & 0 & 0 \\
Partners of & 9 & 0 & 0 & 1 & 0 & 0 &0\\
Women & 10 & 0 & 1 & 0 & 0 & 0 & 0 \\ 
 Participants& 11 & 0 & 0 & 0 & 1 & 0 & 0 \\ 
 & 12 & 0 &0 & 0 & 0 &1& 1 \\ 
 & 13 & 0 & 1 & 0 & 0 & 0&0 \\ 
 & 16 & 1 & 0 & 0 & 0& 0 & 1\\
& 21 & 0 & 0 & 1 & 0 & 0 & 0 \\
\hline
\end{tabular}
\label{bjd1}
\end{center}
\end{table}

\begin{table}[htp]

\centering
\caption{The joint-degree probability distribution for heterosexual partnerships used in the computer simulations. Men and women are assumed to have fewer than $21$ and $6$ partners, respectively. The entry in the $i^{th}$ row and $j^{th}$ column is the fraction of partnership (edges) between men who have $i$ partners (degree $i$) and women who have $j$ partners (degree $j$).}
\begin{tabular}{ccccccc}
\toprule[1.5pt]
\diaghead{\theadfont Diag ColumnmnHead II}%
{\textbf{Degree}\\ \textbf{of men}}{\textbf{Degree of} \\
\textbf{ women}}& $1$ & $2$ & $3$ & $4$ & $5$ & $6$ \\[0.50ex]
\hline

$1$ & $0.3775$ & $0$ & $0$ & $0$ & $0$ & $0$ \\[0.50ex]
\hline
$2$ & $0.1880$ & $0.0205$ & $0$ & $0$ & $0$ & $0$ \\[0.50ex]
\hline
$3$ & $0.0988$ & $0.0396$ & $0.0162$ & $0.0033$ & $0$ & $0$ \\[0.50ex]
\hline 
$4 $ & $0.0833$ & $0.0734$ & $0.0285$ & $0.0087$ & $0.0034$ & $0.0010$ \\[0.50ex]
\hline
$5 $ & $0.0313$ & $0.0264$ & $0.0162$ & $0.0083$ & $0.0037$ & $0.0014$ \\[0.50ex]
\hline 
$6$ & $0.0207$ & $0.0209$ & $0.0134$ & $0.0066$ & $0.0025$ & $0.0003$ \\[0.50ex]
\hline $7 $ & $0.0150$ & $0.0167$ & $0.0107$ & $0.0046$ & $0.0008$ & $0$ \\[0.50ex]
\hline 
$8 $ & $0.0116$ & $0.0135$ & $0.0082$ & $0.0026$ & $0$ & $0$ \\[0.50ex]
\hline
$9 $ & $0.0093$ & $0.0110$ & $0.0061$ & $0.0008$ & $0$ & $0$ \\[0.50ex]
\hline
$10 $ & $0.0075$ & $0.0090$ & $0.0043$ & $0$ & $0$ & $0$ \\[0.50ex]
\hline
$11 $ & $0.0061$ & $0.0074$ & $0.0027$ & $0$ & $0$ & $0$ \\[0.50ex]
\hline
$12 $ & $0.0050$ & $0.0062$ & $0.0014$ & $0$ & $0$ & $0$ \\[0.50ex]
\hline
$13 $ & $0.0042$ & $0.0053$ & $0.0005$ & $0$ & $0$ & $0$ \\[0.50ex]
\hline
$14 $ & $0.0034$ & $0.0045$ & $0$ & $0$ & $0$ & $0$ \\[0.50ex]
\hline
$15 $ & $0.0027$ & $0.0037$ & $0$ & $0$ & $0$ & $0$ \\[0.50ex]
\hline
$16 $ & $0.0021$ & $0.0031$ & $0$ & $0$ & $0$ & $0$ \\[0.50ex]
\hline $17 $ & $0.0016$ & $0.0027$ & $0$ & $0$ & $0$ & $0$ \\[0.50ex]
\hline
$18 $ & $0.0012$ & $0.0023$ & $0$ & $0$ & $0$ & $0$ \\[0.50ex]
\hline 
$19 $ & $0.0009$ & $0.0020$ & $0$ & $0$ & $0$ & $0$ \\[0.50ex]
\hline 
$20 $ & $0.0007$ & $0.0017$ & $0$ & $0$ & $0$ & $0$ \\[0.50ex]
\hline 
$21$ & $0.0059$ & $0.0070$ & $0.020$ & $0$ & $0$ & $0$ \\[0.50ex]
\bottomrule[1.5pt]
\end{tabular}
\label{bjd2}
\end{table}

\clearpage
\newpage

The Fig \ref{net} is a visualization of a random sexual network of $5000$ individuals generated using data in Tables \ref{bjd1} and \ref{bjd2} and the algorithm in \cite{boroojeni2017generating}.

\begin{figure}[htp]
\centering
\includegraphics[scale=.4]{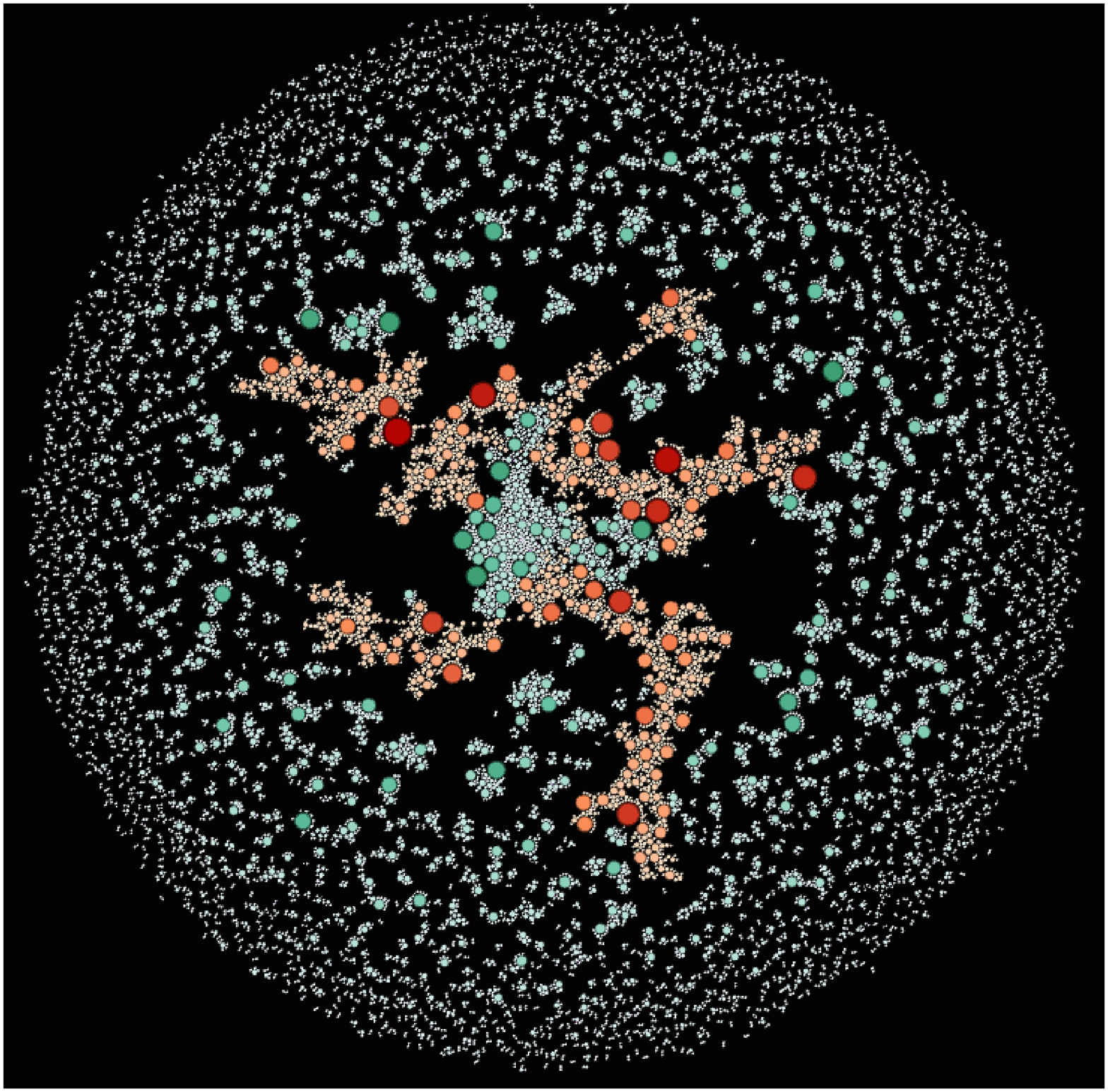}
\caption{\textbf{Visualization of generated heterosexual network of \pmb{$5000$} individuals:} Orange family color is the giant component and green parts are the other components of the network. The size of each node is related to its degree, more partners, bigger the node.}
\label{net}
\end{figure}

In the Table \ref{ci}, we report confidence intervals of prevalence at quasi-stationary state for all population with respect to some of the screening parameter values, some of the partner notification parameter values when $\theta_t=1$, and some of the rescreening parameter values. 
\begin{table}[htp]
\begin{adjustwidth}{-0in}{0in} 
\centering
\caption{The $95\%$ confidence intervals for the fraction of men screened per year ($\sigma_y^m$), the fraction of partners notified when all follow treatment ($\theta_n$ when $\theta_t=1$), and the fraction of people return for screening ($\sigma_r$).
}
\resizebox{\columnwidth}{!}{\begin{tabular}{p{1in}cccccc}
\toprule[1.5pt]
\textbf{Parameter Value} & $0$ & $0.2$ & $0.4$ & $0.6$ & $0.8$ & $1$ \\[0.50ex]
\hline

$\pmb {95\%}$ \textbf{CI for} $\pmb{\sigma_y^m}$ & $[ 0.1256, 0.1294]$ & $[0.1192,0.1227]$ & $[ 0.1038,0.1092]$ & $[0.0983,0.1076]
$ & $[0.0844,0.0919]$ & $[0.0768,0.0878]$ \\[0.50ex]
\hline
$\pmb {95\%}$ \textbf{CI for} $\pmb{\theta_n}$ when $\pmb{\theta_t}=1$ & $[ 0.1307,0.1424]$ & $[ 0.1319,0.1383]$ & $[0.1181,0.1222]$ & $[ 0.1041,0.1075]
$ & $[0.0837,0.0895]$ & $[0.0726,0.0766]$ \\[0.50ex]
\hline
$\pmb {95\%}$ \textbf{CI for} $\pmb{\sigma_r}$ & $[ 0.1270,0.1306]$ & $[ 0.1225,0.1265]$ & $[0.1186,0.1230]$ & $[ 0.1181,0.1224 ]$ & $[0.1052 ,0.1146]$&$[ 0.1097,0.1145]$ \\[0.50ex]
\bottomrule[1.5pt]
\end{tabular}}
\label{ci}
\end{adjustwidth}
\end{table}
\begin{figure}[htp]
\centering
\includegraphics[scale=.33]{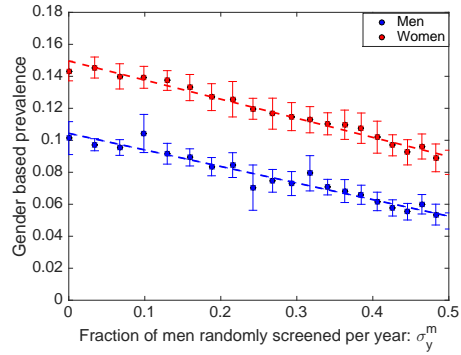}
\caption{\textbf{Prevalence of men and women versus fraction of men screened each year.}}
\label{g_scr}
\end{figure}
\begin{figure}[htp]
\centering
\includegraphics[scale=.33]{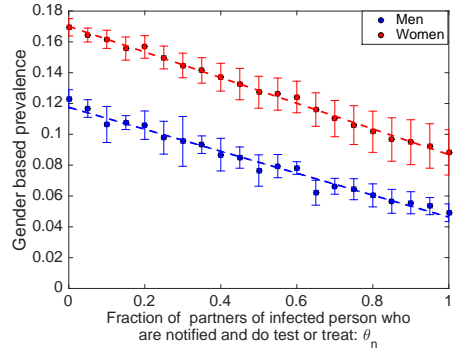}
\caption{{\textbf{Prevalence of men and women versus the fraction of the partners of treated people who are treated after being notified.}}}
\label{g_pt}
\end{figure}
\begin{figure}[htp]
\centering
\includegraphics[scale=.33]{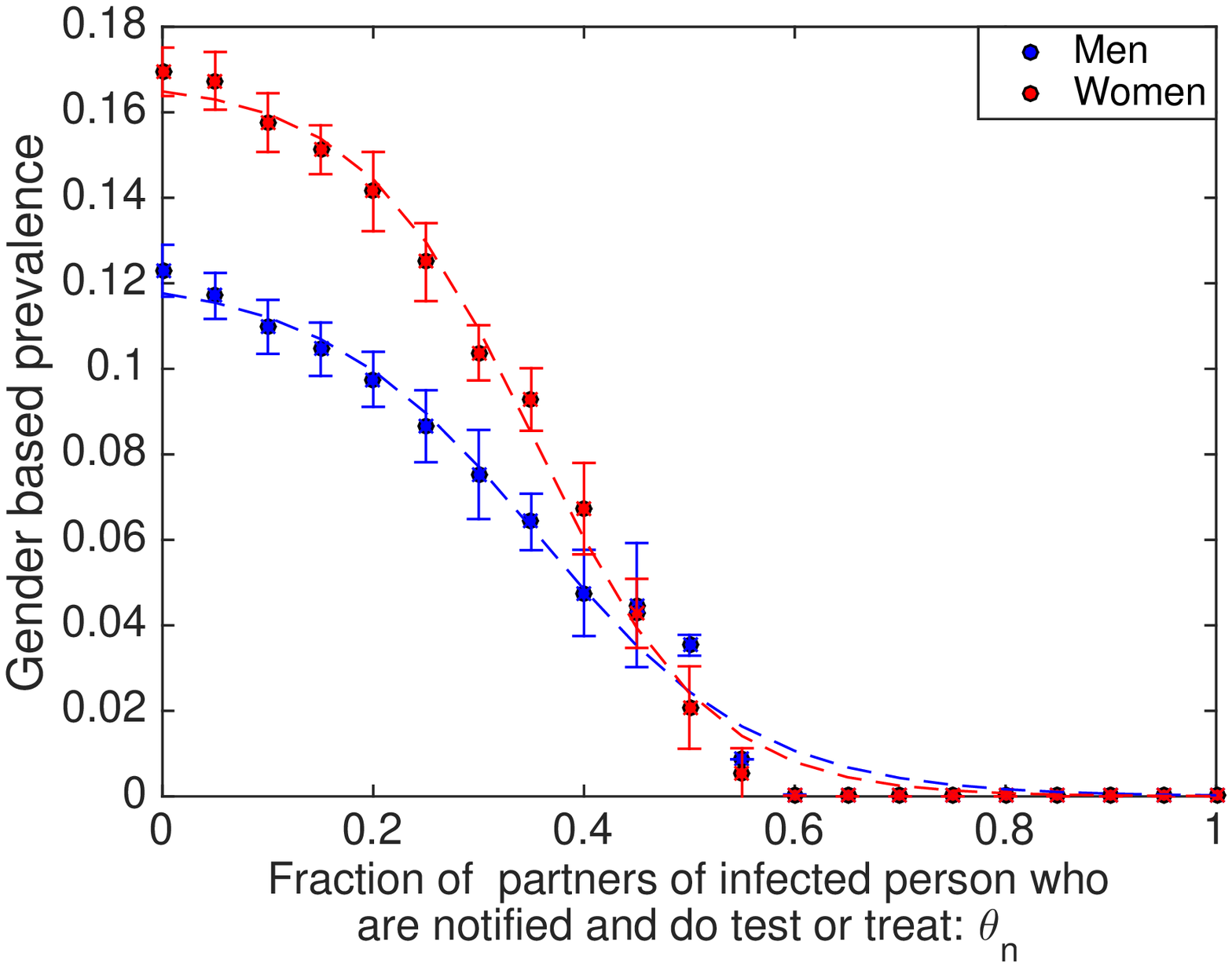}
\caption{\textbf{\textbf{Prevalence of men and women versus the fraction of the partners of treated people who are tested before [H] treatment.}}}
\label{g_ptt}
\end{figure}

Fig \ref{g_scr} shows the impact of men screening on prevalence among men and women separately for the case where the network is static. We observe that screening intervention is insensitive to how often the casual partners change.
Fig. \ref{g_pt} shows the impact of partner treatment on prevalence among men and women separately for the case where the network is static. We observe that partner treatment intervention is insensitive to partner change in time.
When the network is static, the logistic curve steepens to become a threshold condition (tipping point) where at $\theta_n\approx 0.4$ the epidemic can be brought under control.
The tipping point results from the partner screening percolating through the sexual network to identify the infected individuals, Fig \ref{g_ptt}.
\end{document}